\documentclass[twocolumn,showpacs,preprintnumbers,amsmath,amssymb]{revtex4}

\usepackage{graphicx,psfrag}
\usepackage{dcolumn}
\usepackage{bm}

\newcommand{\bea}{\begin{eqnarray}}
\newcommand{\eea}{\end{eqnarray}}

\begin{document}

\preprint{TU--642}
\preprint{Dec.\ 2001}

\title{Corrections to quarkonium \boldmath{$1S$} 
energy level at \boldmath{${\cal O}(\alpha_S^4 m)$} from \\
non-instantaneous Coulomb vacuum polarization }

\author{Y. Kiyo}
\author{Y. Sumino}
\affiliation{Department of Physics, Tohoku University, Sendai, 980-8578 Japan  }

\date{\today}

\begin{abstract}
We calculate ${\cal O}(\alpha_S^4 m)$ and 
${\cal O}(\alpha_S^4 m \log \alpha_S)$
corrections to the quarkonium $1S$ energy level analytically, 
which have been overlooked in recent studies.
These are the contributions from 
one Coulomb-gluon exchange
with the 1-loop vacuum polarization insertion.
A part of the corrections was computed numerically some time ago;
we correct its error.
\end{abstract}



\maketitle


Heavy quarkonium systems, such as bottomonium and charmonium, 
provide an important 
testing ground for studies on physics and dynamics of
QCD boundstates using non-relativistic boundstate theory based on
perturbative QCD.
Recently accuracy of the theoretical predictions has improved dramatically.
This owes to the progress in computations of higher-order 
perturbative corrections
and the discovery \cite{ren-cancel} of
cancellation of leading renormalons in these systems.
So far, impacts in related fields have been as follows.
(1) The mass of the bottom quark in the $\overline{\rm MS}$ scheme
has been determined accurately from the $\Upsilon$ spectrum \cite{b-mass}.
It has various important applications in $B$ physics.
Also it posed tight constraints on
some models beyond the Standard Model such as super-symmetric 
Grand Unified Theories \cite{su5gut}.
(2) The bottomonium spectrum is reproduced reasonably well 
by perturbative QCD up to the $n=3$ levels.
A new physical picture on the composition of the bottomonium masses has
been proposed on the basis of the calculation \cite{BraSuminoVai}.
(3) After incorporating the renormalon cancellation,
the perturbative static QCD potential is shown to agree well with 
potentials of phenomenological models and with the QCD potential
computed by lattice simulations \cite{SuminoReck}.

Despite of these developments,
in the context of non-relativistic boundstate theory,
there have been no systematic methods for identifying higher-order
corrections to physical quantities of heavy quarkonia
(energy levels, decay rates, etc.)\
in the non-relativistic expansion in $1/c$
($c$ is the speed of light). 
Rather, experts have identified separate contributions from inspections.
A first attempt to compute ${\cal O}(1/c^2)$ corrections to the
quarkonium energy levels was given in \cite{duncan},
although its main results contain errors.\footnote{
As for the main results $\Delta^v M_{nl}$ and $\Delta^{\rm se} M_{nl}$ 
of \cite{duncan}, the sign of the former correction is wrong;
we discuss the error of the latter correction
below.
}
In terms of the (effective) non-relativistic Hamiltonian for 
quarkonium systems, different terms up to ${\cal O}(1/c^2)$
have been identified as follows.
The part which is identical with that for QED boundstates has been known
for a long time (the Breit Hamiltonian).
The 1-loop correction to the static QCD potential was computed
in \cite{FischBill}; the 2-loop correction was calculated in \cite{peter}.
The $1/r^2$-type potential unique to non-abelian gauge theory
was derived in \cite{duncan,gupta}.
The analytic expression of the quarkonium energy levels have
been derived from the above Hamiltonian in \cite{pineda98}.
It has been considered that at this stage
all the corrections up to 
${\cal O}(1/c^2) = {\cal O}(\alpha_S^4 m)$ 
to the energy levels 
have been identified and calculated.
($\alpha_S$ is the coupling constant of QCD.)
However, the ${\cal O}(\alpha_S^4 m)$ correction
from the non-instantaneous part of
the gluon vacuum polarization computed in \cite{duncan}
has not been taken into account in all the studies which appeared
after that computation.
It seems that this part of the results of \cite{duncan} has
never been taken seriously and has been forgotten largely.
For instance, \cite{km} states incorrectly that contributions
from the non-instantaneous
part of the gluon vacuum polarization are at most
${\cal O}( \alpha_S^5 m)$.

Recently many efforts have been devoted
for identifying efficiently and systematically
higher-order corrections in non-relativistic boundstate theory.
For this purpose,
people have constructed effective theories such as non-relativistic
QCD (NRQCD) \cite{NRQFT}
and potential-NRQCD (pNRQCD) \cite{pNRQCD}, or used the asymptotic
expansion of Feynman diagrams in $1/c$ (threshold expansion
technique) \cite{threshold-exp}.
As for the quarkonium energy levels,
these formalisms  provided more solid bases to the consensus
that all the ${\cal O}(1/c^2)$ corrections are taken into account.
People have begun calculations of some of the corrections at
${\cal O}(1/c^3)$ and beyond \cite{n3lo,KiyoSumino00,ren-imp}.
There seems to be a tendency in the field that (in particular) the
pNRQCD formalism is considered to be
already well established for organizing and identifying
higher-order corrections systematically.
Nevertheless, the ${\cal O}(1/c^2)$ correction mentioned above 
(and given in a more complete form in this paper)
seems to be overlooked in this formalism, at least in its present
usage.
To our knowledge, up to date there exists no method which enables
identification of all the higher-order corrections systematically.

\begin{figure}
\psfrag{kx}{$k$}
\psfrag{Pi}{\hspace{-2mm}$\Pi(k)$}
\psfrag{Phi}{$\phi_{C,1S}$}
\psfrag{PhiBar}{\hspace{-5mm}$\phi^\ast_{C,1S}$}
\psfrag{Q}{$Q$}
\psfrag{QBar}{$\bar{Q}$}
\psfrag{gC}{$g_C$}
\includegraphics[width=3.9cm]{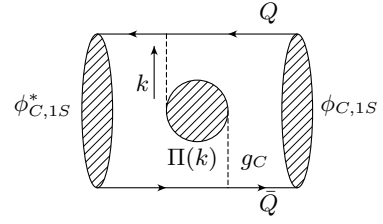}%
\caption{
The correction to the quarkonium potential energy induced by the
1-loop vacuum polarization insertion to the Coulomb gluon propagator.
\label{fig_E1S}}
\end{figure}

In \cite{KiyoSumino01}
we found a correction to the $1S$ energy level at 
${\cal O}( \alpha_S^4 m \log \alpha_S)$ 
(within the approximation of
a bubble-chain resummation).
The correction originates from the ultra-soft
region 
$k_0, |\vec{k}| \sim \alpha_S^2 m$ of the momentum 
of the gluon exchanged between
quark ($Q$) and antiquark ($\bar{Q}$).
From further inspections, we find corrections
to the energy level at ${\cal O}( \alpha_S^4 m \log \alpha_S)$ and
${\cal O}( \alpha_S^4 m )$ originating from the ultra-soft and
the soft momentum region $k_0, |\vec{k}| \sim \alpha_S m$.
Both of these corrections stem from the non-instantaneous nature of
the gluon vacuum polarization.
In this paper we compute these corrections 
to the energy level of the quarkonium $1S$ state
analytically in Coulomb gauge (see Fig.~\ref{fig_E1S}).
We assume that $Q$ and $\bar{Q}$ have equal masses.
The gluonic contribution
corresponds to the correction $\Delta^{\rm se} M_{1S}$ computed
numerically in \cite{duncan}.

Following \cite{KiyoSumino01}, our computation of the perturbative 
corrections is based on the Bethe-Salpeter 
(BS) formalism \cite{BodwinYennie}.
We work in Coulomb gauge, since severe gauge cancellations occur
in other gauges \cite{love}.
(Most computations in non-relativistic boundstate problems
have been carried out in Coulomb gauge conventionally, 
except where gauge-independent
contributions have been computed.)
According to the BS formalism, the potential energy 
$E_{\rm pot}$ 
and the self-energy contributions $E_{\rm SE}$ in the
total energy of the boundstate can be expressed,
in the rest frame of the boundstate, as
\begin{eqnarray}
&&
E_{\rm pot}
=
\frac{i}{2M} \,
\left(
\overline{\chi} \cdot K \cdot \chi 
\right)
~,
\\
&&
E_{\rm SE}
=
M -
\frac{i}{2M} \,
\left(
\overline{\chi} \cdot  \left[
(S_{F,Q})^{-1}(S_{F,\bar{Q}})^{-1} \right]
\cdot \chi 
\right)
~.
\end{eqnarray}
Here, $\chi,\overline{\chi}$ and $K$ denote the BS wave functions
and the BS-kernel, respectively; 
$M$ is the boundstate mass (total energy);
$S_F$ represents the full propagator of $Q$ or $\bar{Q}$.
We set the momentum of the center of gravity as
$(M,\vec{0})$.
The dot ($\cdot$) represents contraction of spinor indices
and an integral over the relative momentum 
between $Q$ and $\bar{Q}$.
Diagrammatically,
$E_{\rm pot}$ ($E_{\rm SE}$) represents the contributions from the diagrams
where the $Q$ and $\bar{Q}$ lines
are connected (disconnected), and
$E_{\rm pot}+E_{\rm SE}=M$.

We compute the contribution to the potential energy
from the 1-loop vacuum polarization of the Coulomb gluon $\Pi (k)$:
\bea
&&
\delta E_{{\rm pot}}^{(1S)} =
\Bigl\langle \, \Pi (k) \, \Bigr\rangle_{1S} 
\nonumber \\ &&
\equiv
\frac{i}{2M^{(0)}_{1S}} \,
\int \! \frac{d^4p}{(2\pi)^4}
\int \! \frac{d^4k}{(2\pi)^4}
\, \, 
\nonumber \\ &&
~~~~~~~~~~~~~~
\times
\overline{\chi}^{(0)}_{1S}(p) 
\, K_{C}(k) \, \Pi(k) \, \chi^{(0)}_{1S}(p+k) .
\label{deltaEpot}
\eea
This is  shown diagrammatically in Fig.~\ref{fig_E1S}.
We assume that the quarks in the vacuum polarization are massless
and compute $\delta E_{{\rm pot}}^{(1S)} $
in a series expansion in $\alpha_S$
up to ${\cal O}( \alpha_S^4 m )$ and ${\cal O}( \alpha_S^4 m \log \alpha_S )$.
In Eq.~(\ref{deltaEpot}),
$
K_C =
-i C_F ({4 \pi\alpha_S}/{|\vec{k}|^2}) 
( \gamma^0 \! \otimes \! \gamma^0 )
$ 
is the kernel of one Coulomb-gluon exchange at tree level.
Throughout this paper
$\alpha_S\equiv\alpha_S(\mu)$ denotes the strong coupling constant 
defined in the
$\overline{\rm MS}$ scheme with $n_l$ active flavors, where
$\mu$ is the renormalization scale;
$C_F $ is the Casimir operator of the fundamental representation of
the color $SU(3)$ group.
The BS wave functions in the leading-order of $1/c$-expansion are given by
\bea
&&
\chi^{(0)}_{1S} (p)=
[ {\cal D}(p) + {\cal D}(-p) ] \,
\sqrt{2M^{(0)}_{1S}} \, \phi_{C,1S}(\vec{p}) \, \lambda  ,
\\ &&
\overline{\chi}^{(0)}_{1S} (p)=
[ {\cal D}(p) + {\cal D}(-p) ] \,
\sqrt{2M^{(0)}_{1S}} \, \phi_{C,1S}^*(\vec{p}) \, \lambda^\dagger ,
\\ &&
{\cal D}(p) = 
\frac{i}{{M^{(0)}_{1S}}/{2}+p_0-m-{\vec{p}^{\,2}}/{2m}+i0} .
\label{chi0}
\eea
Here, $m$ denotes the pole mass of $Q$ and $\bar{Q}$.
$ \phi_{C,1S}(\vec{p})$ and $M^{(0)}_{1S}$ denote the
$1S$ Coulomb wave function and its energy level, respectively:
\bea
&&
\phi_{C,1S}(\vec{p})
=
\frac{ \sqrt{2\pi} \left(C_F \alpha_S m \right)^{ \frac{5}{2}}
    }
{ \left[ \vec{p}^{\,\, 2} + \left( C_F \alpha_S m/2\right)^2 \right]^2}
~,
\\ &&
M^{(0)}_{1S} = 2 \, m - \frac{\bigl( C_F \alpha_S \bigr)^2}{4} \, m .
\eea
$\lambda$ represents the spin of the boundstate,
\bea
&&
\lambda_{J}
=
\frac{1+\gamma^0}{2}\, 
\sigma_J
\frac{1-\gamma^0}{2} ,
\eea
where $\sigma_J={\gamma_5}/{\sqrt{2}}$ for $J=0$ and
$\sigma_J={\varepsilon ~\hspace{-8pt}\slash}/{\sqrt{2}}$
for $J=1$ and polarization vector $\varepsilon$.

\begin{figure}
\psfrag{figA}{(a)}
\psfrag{figB}{(b)}
\psfrag{figC}{(c)}
\psfrag{figD}{(d)}
\psfrag{fl}{$q$}
\psfrag{gC}{$g_C$}
\psfrag{gT}{$g_T$}
\psfrag{gTT}{$g_T$}
\psfrag{gTTT}{$g_T$}
\psfrag{gTTTT}{$g_T$}
\includegraphics[width=5.5cm]{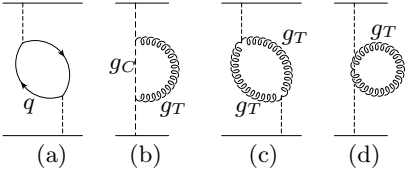}%
\caption{Feynman diagrams for the 1-loop Coulomb vacuum polarization: 
(a) massless quark loop, 
(b) transverse-Coulomb gluon ($g_T$-$g_C$) loop, and 
(c)(d) transverse gluon loops. 
\label{fig_vacuum_pol}}
\end{figure}

First we compute $\Pi (k)$ using dimensional regularization and
in the $\overline{\rm MS}$ renormalization scheme.
(We assume that spacetime contains
one time dimension and $3-2\epsilon$ space dimensions.)
The Feynman diagrams contributing to $\Pi (k)$
are shown in Fig.~\ref{fig_vacuum_pol}.
Since this is a 1-loop calculation, it can be carried out
more or less straightforwardly.
Some useful techniques for calculations in 
Coulomb gauge with dimensional regularization 
can be found in \cite{Coulomb-dim}.
We write
\bea
\Pi(k) = \frac{\alpha_S}{4\pi}\, 
\Bigl[ C_A \, \Pi_{g}(k) + T_R n_l\, \Pi_{q}(k) \Bigr] ,
\eea
where $C_A$ is the Casimir operator of the adjoint representation
and $T_R$ is the trace normalization of the fundamental
representaion.
We obtain the gluon contribution:
\bea
&&
\Pi_{g}(k) =
- \frac{11}{3\,\epsilon} + \Gamma (\epsilon)\, e^{ \gamma_E \epsilon }
\biggl( \frac{\mu^2}{|\vec{k}|^2} \biggr)^\epsilon \, \,
\nonumber \\ &&
\times
\Biggl[ \frac{2 \, \Gamma (1-\epsilon) \Gamma (\frac{3}{2}-\epsilon )}
{\Gamma ( \frac{3}{2}-2\epsilon )}
+ \frac{2 \, (5-4\,\epsilon) \, \Gamma (2-\epsilon)^2 }
{(1-\rho^2)^\epsilon \, \Gamma ( 4-2\epsilon )}
\nonumber \\ &&
+
\epsilon \,
\{ 1 - \rho^2 \, ( 3 - 2\,\epsilon ) \}
\, {\cal K} (\epsilon , \rho )
\Biggr] 
\label{Pi_g-1}
\\ &&
= \frac{11}{3}  \log \biggl( \frac{\mu^2}{|\vec{k}|^2} \biggr)
- \frac{8}{3} \log (1\!-\!\rho^2) + \frac{31}{9}
+ f(\rho)
\nonumber \\ && 
+ \rho^2 \,
[
3  \log (1\!-\!\rho^2)
-12 \log 2 
+ 6 - 3 f(\rho)
] 
+ {\cal O}(\epsilon) ,
\nonumber \\
\label{Pi-g-ana}
\eea
where $\gamma_E = 0.5772...$ is the Euler constant,
$\rho = k_0/|\vec{k}|$, and
\bea
&&
{\cal K}(\epsilon,\rho) = 
\int_0^1 \!\! dx \! \int_0^1 \! \! dy \,
\frac{
x^{1-\epsilon} \, y^{\frac{1}{2}-\epsilon} 
}{
[ \, (1-x\,y)-(1-x)\, \rho^2 \, ]^{1+\epsilon} 
}
,
\label{K}
\\
&&
f(\rho) = \frac{1-\rho^2}{\rho}
\left[ {\rm Li}_2 \biggl( -\frac{1\!-\!\rho}{1\!+\!\rho} \biggr)
+ \frac{1}{4} \log^2 \biggl( \frac{1\!+\!\rho}{1\!-\!\rho} \biggr) 
+ \frac{\pi^2}{12}
\right] 
\nonumber \\ &&
~~~~~~~~~
- 2 \, \log 2 .
\eea
The analytic expression (\ref{Pi-g-ana}) is a new result.\footnote{
A one-parameter integral representation is given in \cite{duncan}.
}
On the other hand, the contribution from a massless quark
is gauge independent and well known:
\bea
\Pi_{q}(k) &=&
\frac{4}{3\,\epsilon} - \Gamma (\epsilon)\, e^{ \gamma_E \epsilon }
\biggl( \frac{\mu^2}{-k^2} \biggr)^\epsilon \, \,
\frac{8 \, \Gamma (2-\epsilon)^2 }{\Gamma ( 4-2\epsilon )}
\label{Pi_q-1}
\\ 
&=& 
- \frac{4}{3} \log \biggl( \frac{\mu^2}{-k^2} \biggr) - \frac{20}{9}
 + {\cal O}(\epsilon) .
\eea
For $k_0^2 > |\vec{k}|^2$ the vacuum polarization $\Pi(k)$ acquires
an imaginary part;
in this case, the usual $+i0$ prescription, 
$k_0^2 \to k_0^2 + i0$, is understood.
If we take the instantaneous limit $k_0 \to 0$ of the Coulomb propagator
$- C_F \, ( 4\pi \alpha_S / |\vec{k}|^{2} ) \,
[ 1 + \Pi(k) ]$,
the 1-loop static QCD potential in momentum space \cite{FischBill}
is correctly reproduced.
It is ensured by the Ward identity in Coulomb gauge \cite{feinberg}.

Next we evaluate the integral (\ref{deltaEpot}).
One may easily verify that the integral is finite
both in ultraviolet and infrared regions.
Using the expressions (\ref{Pi_g-1}), (\ref{K}) and (\ref{Pi_q-1})
for $\Pi(k)$, 
we may take advantage of the functions $J(\epsilon)$ and
$I(\epsilon , \Delta)$ analyzed in \cite{KiyoSumino01}.
These functions are defined similarly to
$\delta E_{{\rm pot}}^{(1S)}$, where
$\Pi(k)$ is replaced by $|\vec{k}|^{-2\epsilon}$ 
or by
$(-\xi^2 k_0^2 + |\vec{k}|^2 - i 0)^{-\epsilon}$ in Eq.~(\ref{deltaEpot}):
\bea
&&
\left\langle |\vec{k}|^{-2\epsilon}
\right\rangle_{1S}
= 
- \, \frac{\bigl( C_F \alpha_S \bigr)^2 m}{2} 
\left( {C_F \alpha_S m} \right)^{-2\epsilon} 
J(\epsilon) ,
\nonumber \\ &&
\left\langle 
( {-\xi^2 k_0^2 + |\vec{k}|^2 - i 0} )^{-\epsilon }
\, \right\rangle_{1S}
\nonumber \\ &&
~~~~~~~
= 
- \, \frac{\bigl( C_F \alpha_S \bigr)^2 m}{2} 
\left( {C_F \alpha_S m} \right)^{-2\epsilon} 
{I}(\epsilon, {\xi}\Delta) .
\eea
Here, $\Delta = C_F \alpha_S/2$ represents the ratio of the
Coulomb binding energy and the Bohr scale.
Then $\delta E_{{\rm pot}}^{(1S)}$ can be expressed 
in terms of $I(\epsilon,\Delta)$ and $J(\epsilon)$ plus
an integral over $x$ and $y$ of terms involving
$I(\epsilon,\sqrt{\frac{1-x}{1-xy}}\Delta)$ and its derivative
with respect to $\Delta$.
In order to obtain the series expansion of 
$\delta E_{{\rm pot}}^{(1S)}$ in $\alpha_S$, 
it suffices to replace $I(\epsilon,\Delta)$
by its asymptotic expansion in $\Delta$ given by
\bea
&&
I(\epsilon,\Delta) =
I^{(0)}_{\rm A}(\epsilon) + \Delta \cdot I^{(1)}_{\rm A}(\epsilon)
+  \Delta^{1-2\epsilon}\cdot I^{(0)}_{\rm NA}(\epsilon)
+ \cdots .
\nonumber \\
\label{asympt-exp}
\eea
Only those terms relevant to our calculation are written explicitly.
$I^{(0)}_{\rm A}$ and $I^{(0)}_{\rm NA}$ were
given in \cite{KiyoSumino01}, while $I^{(1)}_{\rm A}$ is needed additionally:
\bea
I^{(1)}_{\rm A}(\epsilon) = {4}\,{\pi^{-3/2}} \,
\Gamma (1-\epsilon) \, \Gamma ({\textstyle \frac{1}{2} + \epsilon}) 
.
\eea
The integral over $x$ and $y$ can be evaluated analytically.
We obtain
\bea
&&
\delta E_{{\rm pot}}^{(1S)} =
- \, \frac{1}{4} \, \bigl( C_F \alpha_S \bigr)^2 \, m
\nonumber \\ &&
\times
\Biggl[
\Bigl( \frac{\alpha_S}{\pi} \Bigr) 
\Bigl( C_A \, d_g^{(1)} + T_R n_l \, d_q^{(1)} \Bigr)
\nonumber \\ &&
+
\Bigl( \frac{\alpha_S}{\pi} \Bigr)^2 
\Bigl( C_A C_F \, d_g^{(2)} + T_R n_l C_F \, d_q^{(2)} \Bigr)
+ {\cal O}(\alpha_S^3)
\Biggr]
\eea
with
\bea
&&
d_g^{(1)} = \frac{11}{3} \,
\log \biggl( \frac{\mu}{C_F \alpha_S m} \biggr)
+ \frac{97}{18} ,
\\ &&
d_q^{(1)} = 
- \frac{4}{3} \,
\log \biggl( \frac{\mu}{C_F \alpha_S m} \biggr)
- \frac{22}{9} ,
\\ &&
d_g^{(2)} = 
\biggl( \frac{28}{3}-\pi^2 \biggr) \! \log ( C_F \alpha_S )
-10 + \frac{\pi^2}{4} + 7 \zeta_3 ,
\\ &&
d_q^{(2)} = 
- \frac{8}{3} \log ( C_F \alpha_S ) + 4 .
\eea
The ${\cal O}( \alpha_S^3 m )$ terms [$d_g^{(1)}$ and $d_q^{(1)}$]
stem from the static QCD potential and
are well known.
The ${\cal O}( \alpha_S^4 m )$ terms [$d_g^{(2)}$ and $d_q^{(2)}$]
are the new results.
We made a cross check of our results through
an independent computation using a different integral
representation of $\Pi(k)$ and a slightly generalized version of the function
$I(\epsilon,\Delta)$.
We also checked our results by calculating the momentum-space
integral (\ref{deltaEpot}) numerically.

One may use the threshold expansion 
technique \cite{threshold-exp}
in order to clarify from which kinematical regions individual perturbative
corrections originate.
We confirmed that, with respect to the quark-loop contribution, 
the potential region 
($k_0 \sim \alpha_S^2 m$, $|\vec{k}| \sim \alpha_S m$)
accounts for $d_q^{(1)}$;
the soft region induces 
$-(8/3)\log [\mu/(2 C_F \alpha_S m)]$ plus a constant;
the ultra-soft region induces
$+(8/3)\log [\mu/(2 C_F^2 \alpha_S^2 m)]$ plus a constant;
the latter two contributions add up to produce $d_q^{(2)}$.
Similar separate contributions can be identified also for the gluonic
contribution.

As stated, the gluonic contribution $d_g^{(2)}$ was computed numerically
in \cite{duncan}.
Comparing our result with
the numerical results listed in TABLE I of that paper, we find that
their results for the $1S$ state are smaller than ours by about 
factor 6, consistently for all the listed values of $\alpha_S$.
(Note that $\alpha_S$ in \cite{duncan} should read $C_F \alpha_S$
in contemporary notations.)
We have checked that Eq.~(5.13) of that paper is correct.
Therefore, we suspect that some error has occurred in transforming
this equation to the final results
given in the table of that paper.
We are unable to locate the error more precisely, since no details
are provided for this part of the computation.

Let us compare our results with the other ${\cal O}(\alpha_S^4m)$
corrections:
\bea
&&
\delta E_{1S} =
- \, \frac{1}{4}\, \bigl( C_F \alpha_S \bigr)^2 \, m 
\times
\Bigl( \frac{\alpha_S}{\pi} \Bigr)^2 
\nonumber \\ &&
~~~~~~~~~ ~~~~~~~~~
\times
( A_{\rm QCDpot} + A_{1/r^2} + A_{\rm Breit} ) .
\eea
We separate the corrections into 3 parts:
$A_{\rm QCDpot}$ denotes the correction originating from the
static QCD potential;
$A_{1/r^2}$ denotes the correction originating from the
$- C_A C_F \alpha_S^2/(2mr^2)$ potential;
$A_{\rm Breit}$ denotes the correction originating from the
Breit Hamiltonian.
A numerical comparison is given in Table~\ref{tab1}
for $n_l=4$ 
and for the standard values of the color factors
$C_F = 4/3$, $C_A=3$ and $T_R=1/2$.
We see that the corrections
$C_A C_F \, d_g^{(2)}$ and $T_R n_l C_F \, d_q^{(2)}$
are not negligible 
as compared to $A_{1/r^2}$ or $A_{\rm Breit}$,
whereas $A_{\rm QCDpot}$
is an order of magnitude larger 
than the other corrections (for a typical choice of $\mu$)
due to an enhancement by ${\cal O}(\Lambda_{\rm QCD})$ renormalon
\cite{al}.
\begin{table}[t]
\begin{tabular}{l|c}
\hline
$C_A C_F \, d_g^{(2)}$ & $3.5 - 2.1 \, \log (C_F\alpha_S)$
\\
$T_R n_l C_F \, d_q^{(2)}$ & $10.7 - 7.1 \, \log (C_F\alpha_S)$
\\
$A_{\rm QCDpot}$ & $153.6 + 119.1 \, \ell + 52.1 \, \ell^2$
\\
$A_{1/r^2}$ & $39.5$
\\
$A_{\rm Breit}(J=1)$ & $-0.4$
\\
$A_{\rm Breit}(J=0)$ & $23.0$
\\
\hline
\end{tabular}
\caption{ \label{tab1}
A numerical comparison of our results with
the previously known corrections at ${\cal O}(\alpha_S^4 m)$.
We take $n_l=4$.
$\ell = \log[ \mu/(C_F \alpha_S m) ]$.
}
\end{table}

The reason why the corrections calculated here have been overlooked
in the effective theories appears to be
two-fold:
(1) The present power counting schemes cannot specify all diagrams
(and kinematical regions) which contribute
to a given order of $1/c$ expansion.
(2) The effective theories have been matched to  on-shell $Q\bar{Q}$
amplitudes of perturbative QCD;
generally the matching should be performed with off-shell
$Q\bar{Q}$ amplitudes or including on-shell amplitudes with
additional gluons in external lines \cite{gauge-dep}.
We consider that we have not yet understood well
higher-order corrections in the non-relativistic
boundstate theory, especially where the soft and ultra-soft
contributions are involved. 

Our results given here also apply to the 
QED boundstates composed of heavy particles such as
the $\mu^+ \mu^-$ boundstate, after trivial replacements
of the color factors and the coupling constant.


\begin{acknowledgments}
Y. K. was supported by the Japan Society for the Promotion of Science.
\end{acknowledgments}


\end{document}